\documentclass[12pt]{article}

\usepackage{scicite}

\usepackage{times}

\newenvironment{sciabstract}{%
\begin{quote} \bf}
{\end{quote}}

\newcounter{lastnote}
\newenvironment{scilastnote}{%
\setcounter{lastnote}{\value{enumiv}}%
\addtocounter{lastnote}{+1}%
\begin{list}%
{\arabic{lastnote}.}
{\setlength{\leftmargin}{.3in}}
{\setlength{\labelsep}{.5em}}}
{\end{list}}

\usepackage{graphicx}% Include figure files
\usepackage{amsmath}
\usepackage{dcolumn}% Align table columns on decimal point
\usepackage{bm}% bold math
\usepackage{braket} %Dirac notation
\usepackage{bbold}
\usepackage[
 colorlinks=true,
 urlcolor=blue,
 citecolor=blue,
 linkcolor=blue,
 bookmarks=false,
 pdfstartview={FitH},
]{hyperref}% add hypertext capabilities

\title{Chirality density wave of the `hidden order' phase in URu$_2$Si$_2$} 

\author
{H.-H.~Kung$^{1,\ast\ast\ast}$, 
R.~E.~Baumbach$^2$, E.~D.~Bauer$^2$, 
V.~K.~Thorsm{\o}lle$^1$, \\
W.-L.~Zhang$^{1,\dagger}$, K.~Haule$^{1,\ast\ast}$,
J.~A.~Mydosh$^3$ \& G.~Blumberg$^{1,\ast}$\\
\\
\normalsize{$^{1}$Department of Physics \& Astronomy, Rutgers University,
 Piscataway, New Jersey 08854, USA.}\\
\normalsize{$^{2}$Los Alamos National Laboratory, Los Alamos, New Mexico
 87545, USA.}\\
\normalsize{$^{3}$Kamerlingh Onnes Laboratory, Leiden University, 2300 RA
 Leiden, The Netherlands.}\\
\normalsize{$^{\dagger}$On leave from Institute of Physics, Chinese Academy of Sciences, Beijing 100190, China.}\\
\normalsize{To whom correspondence should be addressed, 
e-mails: $^\ast$girsh@physics.rutgers.edu}; \\
\normalsize{$^{\ast\ast}$haule@physics.rutgers.edu}; 
\normalsize{$^{\ast\ast\ast}$skung@physics.rutgers.edu}. 
}
\date{}

\begin{document} 

\baselineskip24pt
\maketitle 
\begin{sciabstract}
A second-order phase transition is associated with emergence of an 
``order parameter'' and a spontaneous symmetry breaking. 
For the heavy fermion superconductor URu$_2$Si$_2$,
the symmetry of the order parameter associated with its ordered phase 
below 17.5\,K 
has remained ambiguous despite 30 years of research,
and hence is called ``hidden order'' (HO).
Here we use polarization resolved Raman spectroscopy to specify the symmetry 
of the low energy excitations above and below the HO transition. 
These excitations involve transitions between interacting heavy uranium 5\textit{f} orbitals, 
responsible for the broken symmetry in the HO phase.
From the symmetry analysis of the collective mode,
we determine that the HO parameter breaks local vertical and diagonal 
reflection symmetries at the uranium sites, 
resulting in crystal field states with distinct chiral properties,
which order to a commensurate chirality density wave ground state.
\end{sciabstract}

\newpage
Electrons occupying 5\textit{f} orbitals often possess dual characters in solids,
partly itinerant and partly localized,
which leads to a rich variety of self-organization at low temperature,
such as magnetism, superconductivity, or even more exotic states\cite{Stewart1984}.
These ordered states are in general characterized by the symmetry they break,
and an order parameter may be constructed to describe the state with reduced symmetry.
In a solid, the order parameter encodes the microscopic interactions 
among electrons that lead to the phase transition. 
In materials containing \textit{f}-electrons, 
exchange interactions of the lanthanide or actinide 
magnetic moments typically generate long-range antiferromagnetic or ferromagnetic order 
at low temperatures, but multipolar ordering
such as quadrupolar, octupolar and hexadecapolar is also possible\cite{Santini2009}.

One particularly interesting example among this class of materials
is the uranium-based inter-metallic compound URu$_2$Si$_2$. 
It displays a non-magnetic second-order phase transition 
into an electronically ordered state at $T_\text{HO}=17.5\,$K, 
which becomes superconducting below 1.5\,K\cite{Palstra1985,Maple1986}.
Despite numerous theoretical proposals to explain the properties
below $T_\text{HO}$ in the past 30 years 
\cite{Santini1994,Chandra2002,Haule2009,Elgazzar2009,Ikeda2012,Chandra2013},
the symmetry and microscopic mechanism for the order parameter remains ambiguous, 
hence the term ``hidden order'' (HO)\cite{Mydosh2011}.
In this ordered state, an energy gap in both spin and charge response have been reported
\cite{Bonn1988,Hall2012,Guo2012,Aynajian2010,Broholm1991,Wiebe2007,Bourdarot2010}.
In addition, an in-gap collective excitation at a commensurate wave vector
has been observed in neutron scattering experiments\cite{Broholm1991,Wiebe2007,Bourdarot2010}.
Recently, four-fold rotational symmetry breaking under an in-plane magnetic field\cite{Okazaki2011} 
and a lattice distortion along the crystallographic \textit{a}-axis\cite{Tonegawa2014} 
has been reported in high quality small crystals.
However, the available experimental works can not yet conclusively determine
the symmetry of the order parameter in the HO phase.

URu$_2$Si$_2$ crystallizes in a body-centered tetragonal structure
belonging to the $\mathbb{D_{4h}}$ point group (space group No.\,139 $I4/mmm$, Fig~1A). 
The uniqueness of URu$_2$Si$_2$ is rooted in the coexistence of the broad conduction bands, 
comprised mostly of Si-\textit{p} and Ru-\textit{d} electronic states,
and more localized U-5\textit{f} orbitals, 
which are in a mixed valent configuration between tetravalent $5f^2$ and trivalent $5f^3$\cite{Jeffries2010}. 
When the temperature is lowered below approximately 70\,K, 
the hybridization with the conduction band allows a small fraction of a U-5\textit{f} electron 
to participate in formation of a narrow quasiparticle band at the Fermi level, 
while the rest of the electron remains better described as localized on the uranium site.

In the dominant atomic configuration\cite{Haule2009,SM}, 
the orbital angular momentums and spins of the two quasi-localized 
U-5\textit{f} electrons add up to total momentum 4$\hbar$, having nine-fold degeneracy.
In the crystal environment of URu$_2$Si$_2$,
these states split into seven energy levels
denoted by irreducible representations of the $\mathbb{D_{4h}}$ group:
5 singlet states $2A_{1g}\oplus A_{2g}\oplus B_{1g}\oplus B_{2g}$
and 2 doublet states $2E_{g}$.
Each irreducible representation possess distinct symmetry properties 
under operations such as reflection, inversion, and rotation.
For example, the $A_{1g}$ states 
are invariant under all symmetry operations of the $\mathbb{D_{4h}}$ group (Fig.~1A),
but the $A_{2g}$ state changes sign under all diagonal and vertical reflections,
and thereby possesses 8 nodes (Fig.~1A).
Most of the measurable physical quantities,
such as density-density and stress tensors, or one particle response functions, 
are symmetric under exchange of \textit{x}- and \textit{y}-axis in tetragonal crystal structure
and therefore do not probe the $A_{2g}$ excitations.
In contrast, these $A_{2g}$ excitations are accessible to Raman spectroscopy
\cite{Shastry1991,Koningstein1968,Rho2004,Cooper1987}.

\begin{figure}
\begin{center}
\vspace{-1 in}
\includegraphics[width=6.3in]{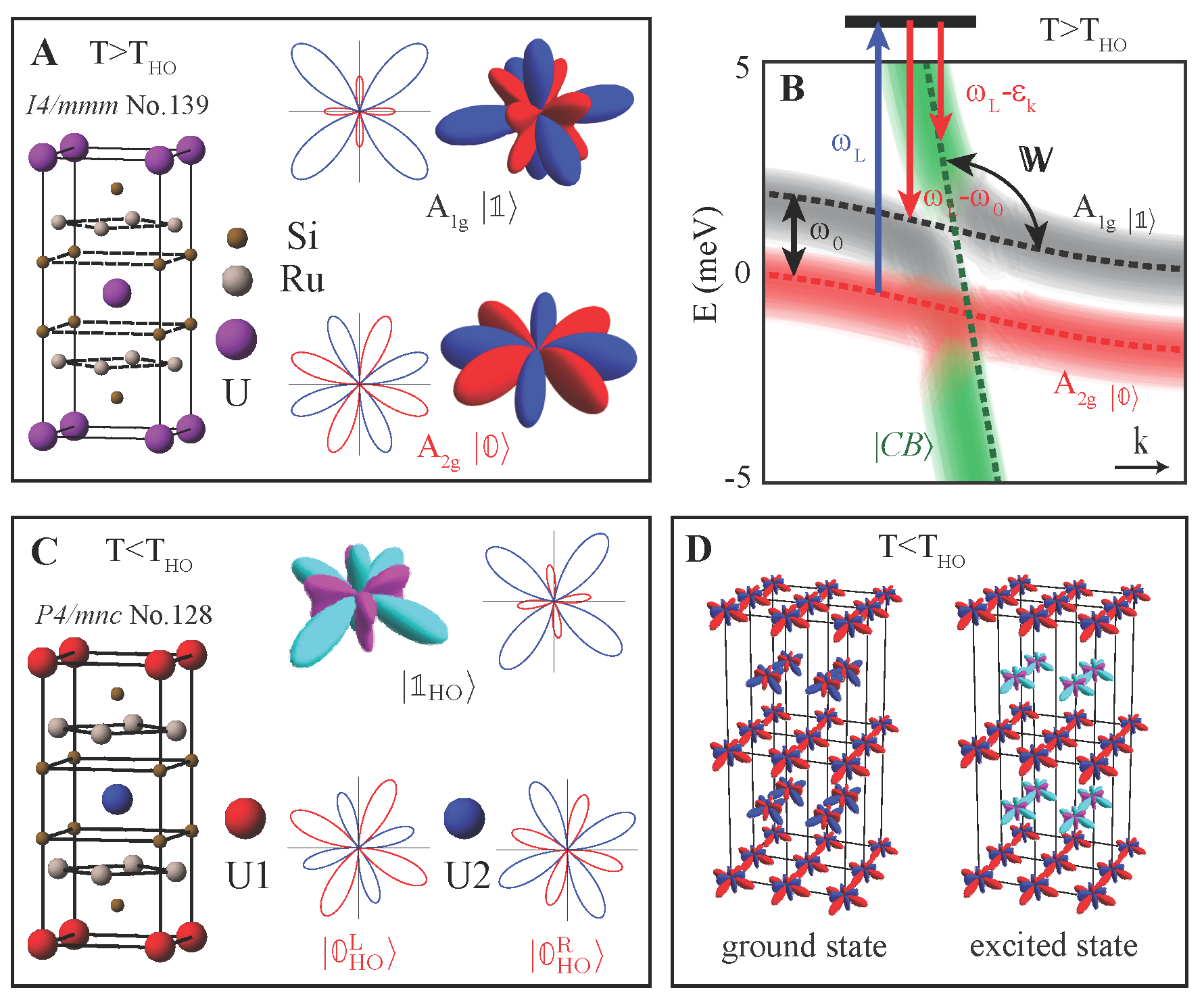}
\end{center}
\caption{
Schematics of the local symmetry and band structure of the quasi-localized states in the minimal model, 
above and below $T_\text{HO}$.
(\textbf{A})
The crystal structure of URu$_2$Si$_2$ in the paramagnetic phase.
Presented in 3D and \textit{xy}-plane cut are wave functions that
show the symmetry of the $A_{2g}$ state $\ket{\mathbb{0}}$ 
and $A_{1g}$ state $\ket{\mathbb{1}}$,
where the positive (negative) amplitude is denoted by red (blue) color.
The $A_{1g}$ state is symmetric with respect to the vertical and diagonal reflections,
while the $A_{2g}$ state is antisymmetric with respect to these reflections. 
(\textbf{B})
Schematic of the band structure of a minimal model in the paramagnetic state.
The green dashed line denotes the conduction band $\ket{CB}$,
the red and black dashed lines denote crystal field states of the U \textit{5f} electrons:
the ground state $\ket{\mathbb{0}}$ and 
the first excited state $\ket{\mathbb{1}}$\cite{SM}. 
A cartoon of the Raman process is shown,
where the blue and red arrows denote the incident and scattered light, respectively.
$\omega_L$ is the incoming photon energy (not in scale),
$\mathbb{W}$ is the hybridization strength between $\ket{\mathbb{1}}$ and $\ket{CB}$,
$\omega_0$ and $\varepsilon_k$
are the resonance energies for $\ket{\mathbb{0}}\rightarrow\ket{\mathbb{1}}$
and $\ket{\mathbb{0}}\rightarrow\ket{CB}$, respectively.
(\textbf{C})
The crystal structure of URu$_2$Si$_2$ in the HO phase,
and wave functions that show the symmetry of the chiral states
$\ket{\mathbb{0}_\text{HO}^\text{L}}$ and 
$\ket{\mathbb{0}_\text{HO}^\text{R}}$,
and the excited state $\ket{\mathbb{1}_\text{HO}}$.
The \textit{left}- and \textit{right}-handed states,
denoted by red and blue atoms, respectively,
are staggered in the lattice as shown.
(\textbf{D}) 
Show schematics of chirality density wave,
where the chiral states are staggered in the lattice (left).
The right figure shows one of the possible excited state of the chirality density wave,
where the chiral state $\ket{\mathbb{0}_\text{HO}^\text{R}}$ at 
lattice site U2 is excited to $\ket{\mathbb{1}_\text{HO}}$.
}
\end{figure}
Raman scattering is an inelastic process which promotes excitations 
of controlled symmetry\cite{SM} (Fig.~1A) defined by the scattering geometries,
namely polarizations of the incident and scattered light\cite{Ovander1960}
(blue and red arrows in Fig.~5).
It enables to separate the spectra of excitations into single symmetry representation\cite{Shastry1991},
such as $A_{1g}$, $A_{2g}$, $B_{1g}$, $B_{1g}$, and $E_g$ in the $\mathbb{D_{4h}}$ group (Fig.~5), 
and thereby classify the symmetry of the collective excitations\cite{SM}.
The temperature evolution of these excitations across
a phase transition provides an unambiguous identification of the broken symmetries.
Unlike most other symmetry sensitive probes requiring external perturbations, 
such as magnetic\cite{Okazaki2011}, electric or strain fields\cite{Riggs2014}, 
the photon field used by Raman probe is weak.
Thus, Raman spectroscopy presents an ideal tool to study the broken symmetries 
across phase transitions without introducing external symmetry breaking perturbations.
\begin{figure}
\begin{center}
\includegraphics[width=6.3in]{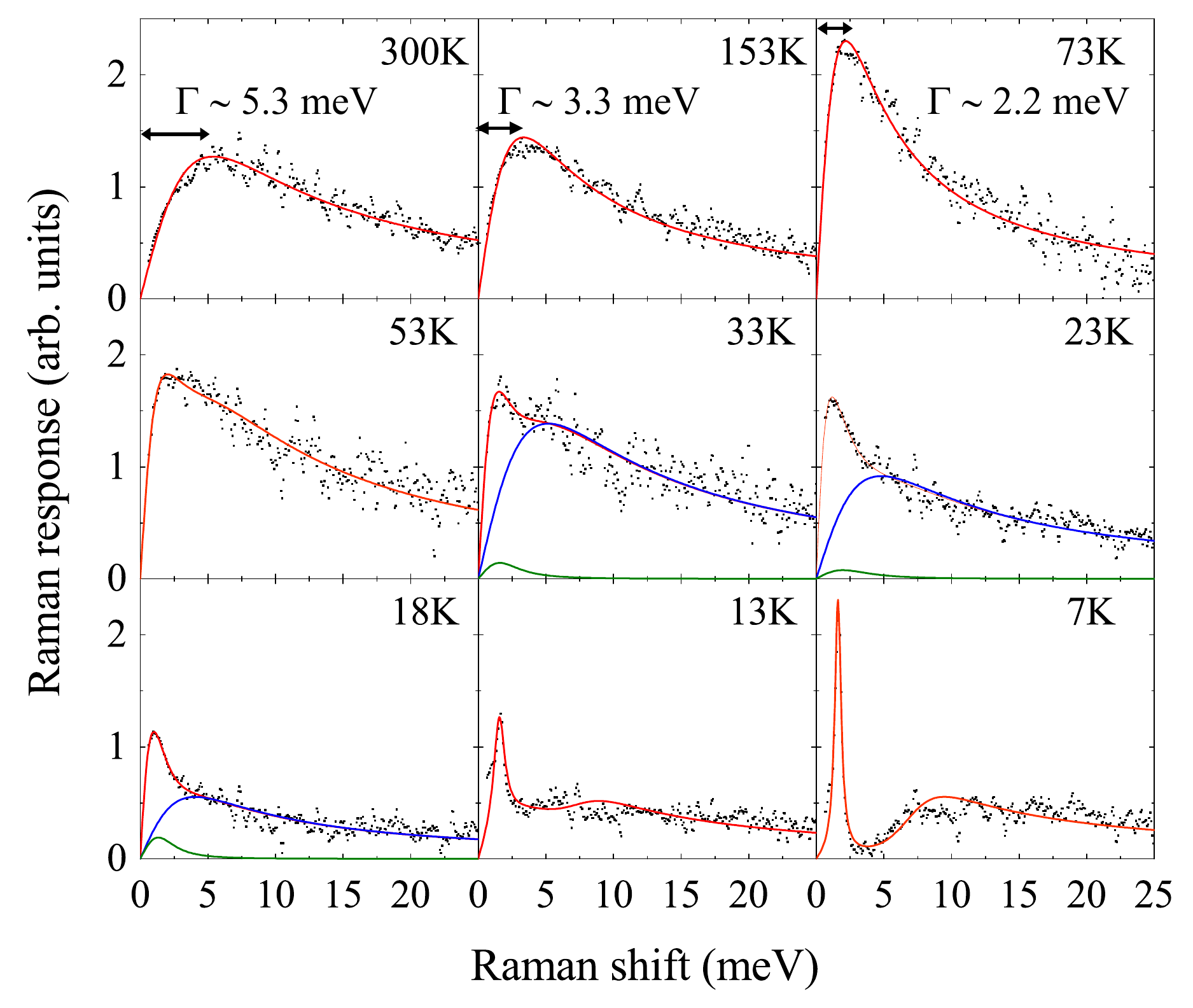}
\end{center}
\caption{
Temperature dependence of the Raman response in A$_{2g}$ symmetry channel (black dots).
The red lines are fitting curves to a Fano model\cite{Klein1983}
within the minimal model described in figure~1B: 
$\chi_{A_{2g}}^{\prime\prime}\!(\omega,T)=-\text{Im}\big[\mathbf{K}^\dagger(\chi_0^{-1}-\mathbb{W})^{-1}\mathbf{K}\big]$~,
where $\mathbf{K}$ is the coupling amplitude to light,
$\mathbb{W}$ is the off-diagonal coupling matrix between the two excitation channels, 
describing the hybridization process shown in figure~1B.
$\chi_0=\text{Diag}\big[(i\omega-\Gamma)^{-1},~[(\omega-\omega_0+i\gamma)^{-1}-(\omega+\omega_0+i\gamma)^{-1}]\big]$
is the unperturbed susceptibility.
The first term is the quasielastic peak, where $\Gamma(T)$ is the peak position.
The second term is a Lorentzian for a resonant transition between the
quasi-localized states $\ket{\mathbb{0}}$ and $\ket{\mathbb{1}}$, with energy
$\omega_0$ and scattering rate $\gamma(T)$.
From 33 to 18\,K, the quasielastic peak and the overdamped 
$\omega_0$ resonance are denoted by blue and green lines, respectively.
The appearance of the maximum at 1\,meV is due to the spectral
weight redistribution resulting from the hybridization coupling $\mathbb{W}$.
Below $T_\text{HO}$ (7 and 13\,K), the spectra show an energy gap
opening and the appearance of another excitation within the gap,
which sharpens dramatically upon cooling.
}
\end{figure}

We employ linearly and circularly polarized light to acquire the
temperature evolution of the Raman response functions in all symmetry channels.
In figure~2 we plot the Raman response in the $A_{2g}$ channel, 
where the most significant temperature dependence was observed.
The Raman response in the paramagnetic state can be described within
a low energy minimal model (illustrated in Fig.~1A, B) 
that contains two singlet states of $A_{2g}$ and $A_{1g}$ symmetries,
split by $\omega_0$,
and a predominantly $A_{1g}$ symmetry conduction band.
In the following, 
we denote the singlet states of $A_{2g}$ and $A_{1g}$ symmetries 
by $\ket{\mathbb{0}}$ and $\ket{\mathbb{1}}$,
as suggested in Ref.~\cite{Haule2009},
and conduction band labeled by $\ket{CB}$.

At high temperatures, the Raman response exhibits quasielastic scattering,
with maximum decreasing from 5\,meV at room temperature 
to 2\,meV at low temperature (Fig.~2).
We interpret these excitations as transitions from the $\ket{\mathbb{0}}$ state into conduction band $\ket{CB}$.
Below 50\,K, a new maximum around 1\,meV develops. 
This feature resembles a Fano-type interference,
where a resonance interacts with the electronic continuum\cite{Klein1983}.
Here, we interpret the two interacting excitations as 
the quasielastic scattering (blue lines in Fig.~2), 
and an overdamped $\omega_0$ resonance between $\ket{\mathbb{0}}$ 
and $\ket{\mathbb{1}}$ states (green lines in Fig.~2).
After turning on the hybridization between $\ket{\mathbb{1}}$ and $\ket{CB}$,
some of the Raman spectral weight is redistributed to lower energy, 
whereby the spectra acquires the observed feature.
Such hybridization tracks the formation of the heavy fermion states in URu$_2$Si$_2$.

Figure~3 displays a comparison between the static Raman
susceptibility $\chi_{A_{2g}}^\prime\!(0)$ (left axis) and the
\textit{c}-axis static magnetic susceptibility $\chi_{c}^m$ (right axis), 
showing that in the paramagnetic phase the responses are proportional to each other.  
This proportionality can be understood by noting that both
susceptibilities probe $A_{2g}$-like excitations, 
which are dominated by transitions from $\ket{\mathbb{0}}$ to conduction band $\ket{CB}$, 
hence in the minimal model of figure~1B, they are proportional to each other.  
The extreme anisotropy of the magnetic susceptibility (Fig.~3) also follows from
this minimal model~\cite{SM}.
\begin{figure}
\begin{center}
\includegraphics[width=4.75in]{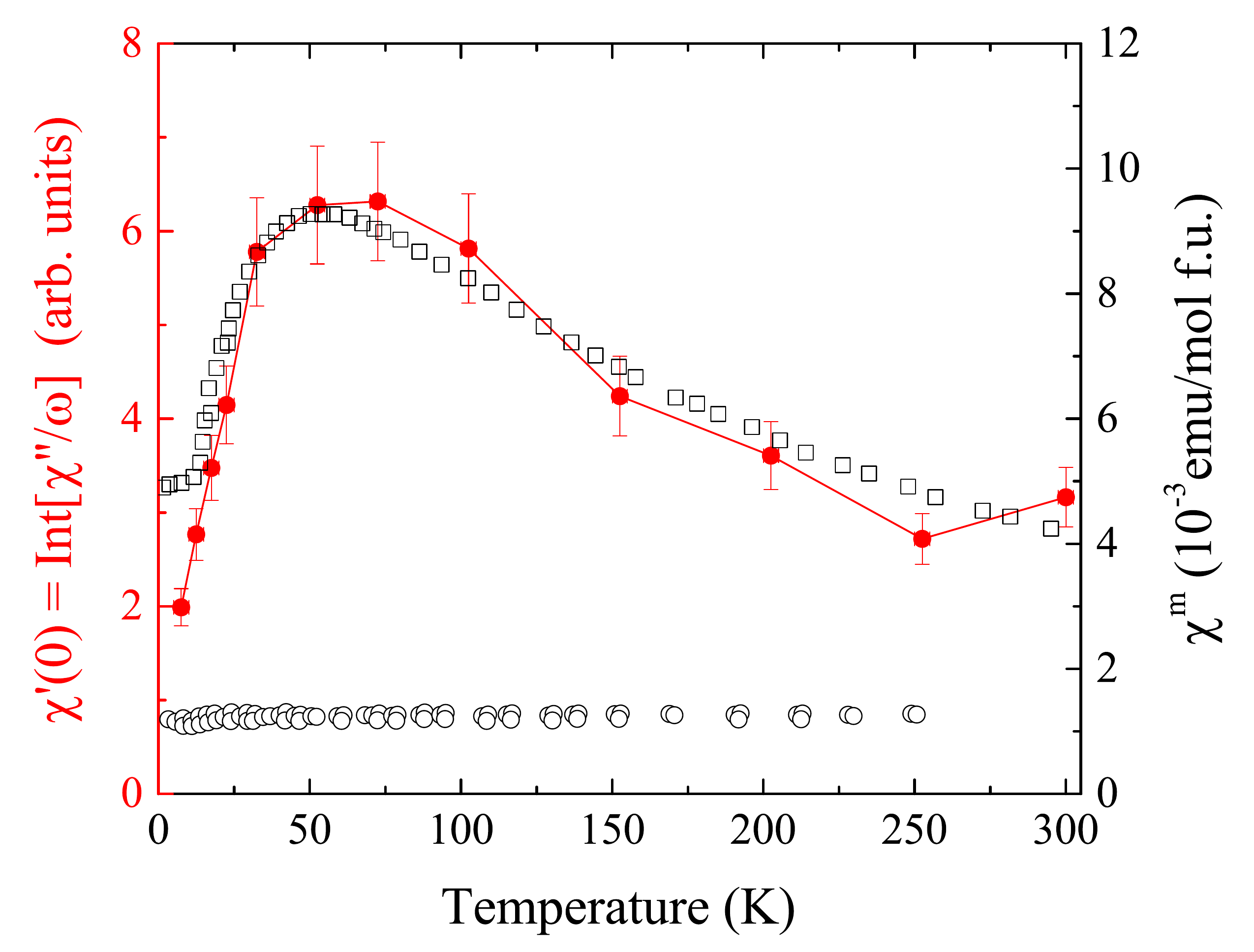}
\end{center}
\caption{
Comparison of static Raman and magnetic susceptibilities.
Temperature dependence of the static Raman susceptibility in $A_{2g}$ channel:
$\chi_{A_{2g}}^\prime\!(0,T)=\frac{2}{\pi}\int_{0}^{\infty}\!\frac{\chi^{\prime\prime}_{A_{2g}}(\omega,T)}{\omega}\,d\omega$
(solid dots),
and the static magnetic susceptibility along \textit{c}- and
\textit{a}-axis from Ref.\cite{Palstra1985} are plotted as open squares and circles, respectively. 
}
\end{figure}

Below 18\,K, the Raman response in the $A_{2g}$ channel (Fig.~2) 
shows the suppression of low energy spectral weight below 6\,meV 
and the emergence of a sharp in-gap mode at 1.6\,meV.
Figure~4 shows the detailed development of these features.
The temperature dependence of the gap qualitatively follows
the gap function expected from a mean-field BCS model (pink line in Fig.~4).
\begin{figure}
\begin{center}
\includegraphics[width=4.75in]{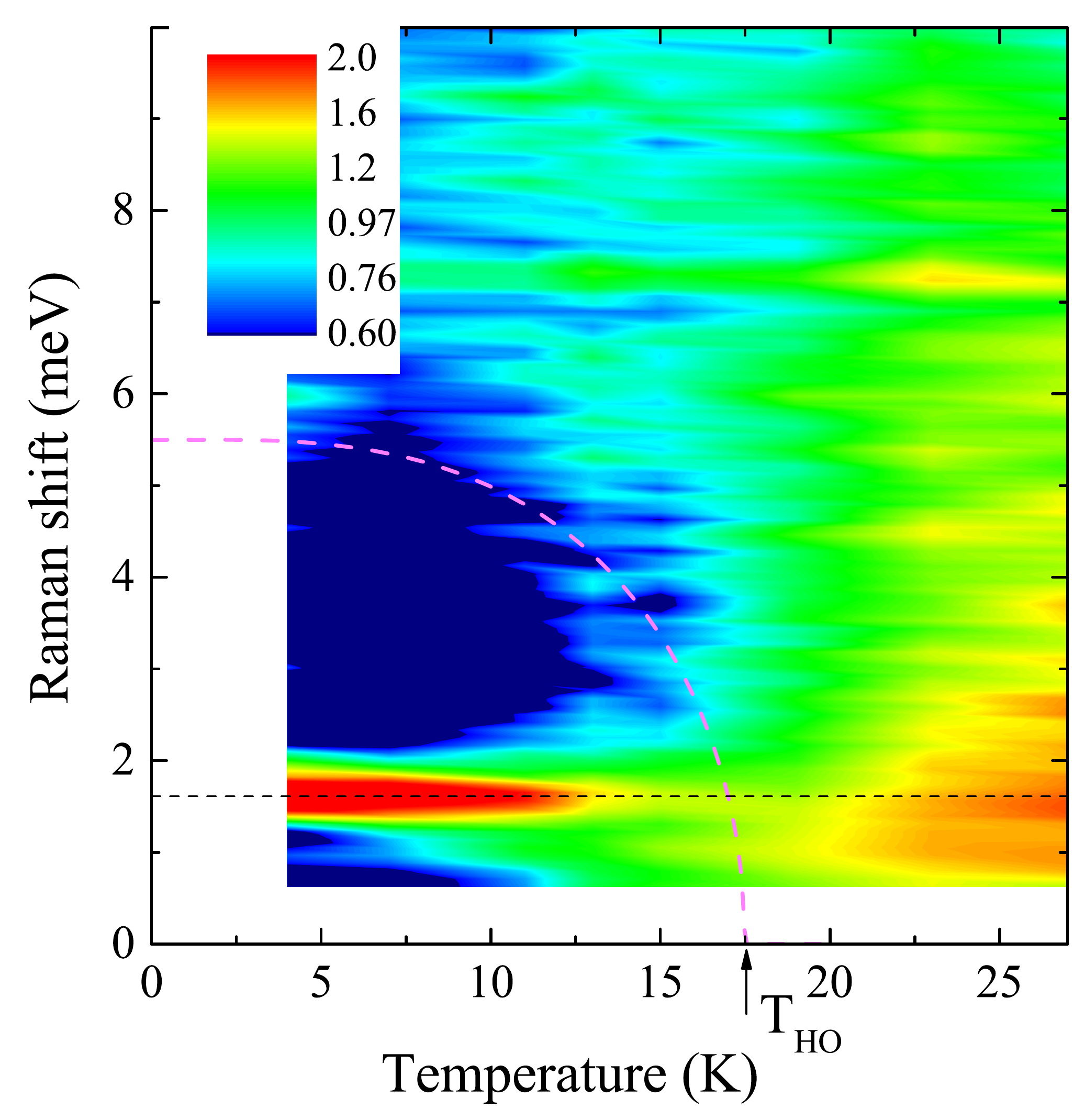}
\end{center}
\caption{
Raman response in the XY scattering geometry as function of 
temperature and Raman shift energy.
The contour plot shows the temperature evolution of the low energy Raman response in the XY scattering geometry.
A sharp excitation at 1.6\,meV (indicated by the black dashed line) emerges below $T_\text{HO}$.
The mode's full width at half maximum decreases on cooling to about 0.3\,meV at 4\,K.
A gap-like suppression develops to a magnitude of about 6\,meV at 4\,K.
The pink dashed line shows the temperature dependence of a gap
expected from a mean-field BCS model with a transition temperature of 17.5\,K.
}
\end{figure}

Having established the Raman response of $A_{2g}$ symmetry and its correspondence 
with the magnetic susceptibility in the paramagnetic state, 
we now present our main results describing the symmetry breaking in the HO state.
Figure~5 shows the Raman response in all six proper scattering geometries at 7\,K.
The intense in-gap mode is observed in all scattering geometries containing $A_{2g}$ symmetry.
The mode can be interpreted as a $\omega_0=1.6$\,meV resonance  
between the $\ket{\mathbb{0}}$ and $\ket{\mathbb{1}}$ quasi-localized states,
which can only appear in the $A_{2g}$ channel of the $\mathbb{D_{4h}}$ group (Fig.~1B).
A weaker intensity is also observed at the same energy 
in XX and X$^\prime$X$^\prime$ geometries commonly containing 
the excitations of the $A_{1g}$ symmetry,
and a much weaker intensity is barely seen within the experimental uncertainty
in RL geometry..
The in-gap mode intensity in the $A_{1g}$ channel 
is about four times weaker than in the $A_{2g}$ channel.

The observation of this intensity ``leakage'' into forbidden 
scattering geometries implies the lowering 
of symmetry in the HO phase,
allowing some of the irreducible representations of $\mathbb{D_{4h}}$ 
point group to mix.
For example, the $\omega_0$ mode intensity ``leakage'' from the $A_{2g}$ into the
$A_{1g}$ channel implies that the irreducible representation $A_{1g}$ and $A_{2g}$  
of the $\mathbb{D_{4h}}$ point group
merge into the $A_g$ representation of the lower group $\mathbb{C_{4h}}$.
This signifies the breaking of the local vertical and diagonal reflection symmetries
at the uranium sites in the HO phase.
Similarly, the tiny intensity leakage into the RL scattering geometry
measure the strength of orthorhombic distortion
due to broken four-fold rotational symmetry.

When the reflection symmetries are broken, 
an $A_{2g}$-like interaction operator $\Psi_\text{HO}\equiv V\ket{\mathbb{1}}\bra{\mathbb{0}}$
mixes the $\ket{\mathbb{0}}$ and $\ket{\mathbb{1}}$ states 
leading to two new local states: 
$\ket{\mathbb{0}_\text{HO}^\text{L}} \approx (1-\frac{V^2}{2\omega_0^2})\ket{\mathbb{0}}-\frac{V}{\omega_0}\ket{\mathbb{1}}$
and $\ket{\mathbb{0}_\text{HO}^\text{R}} \approx (1-\frac{V^2}{2\omega_0^2})\ket{\mathbb{0}}+\frac{V}{\omega_0}\ket{\mathbb{1}}$, 
with $V$ being the interaction strength\cite{Haule2009}.
A pair of such states cannot be transformed into one another 
by any remaining $\mathbb{C_{4h}}$ group operators:
a property known as chirality (or handedness).
The choice of either the left-handed or the right-handed state 
on a given uranium site,
$\ket{\mathbb{0}_\text{HO}^\text{L}}$ or $\ket{\mathbb{0}_\text{HO}^\text{R}}$,
defines the local chirality in the HO phase (Fig.~1C). 
Notice that these two degenerate states both preserve the time reversal symmetry, 
carry no spin and contain the same charge, 
but differ only in handedness.
\begin{figure}
\begin{center}
\includegraphics[width=6.3in]{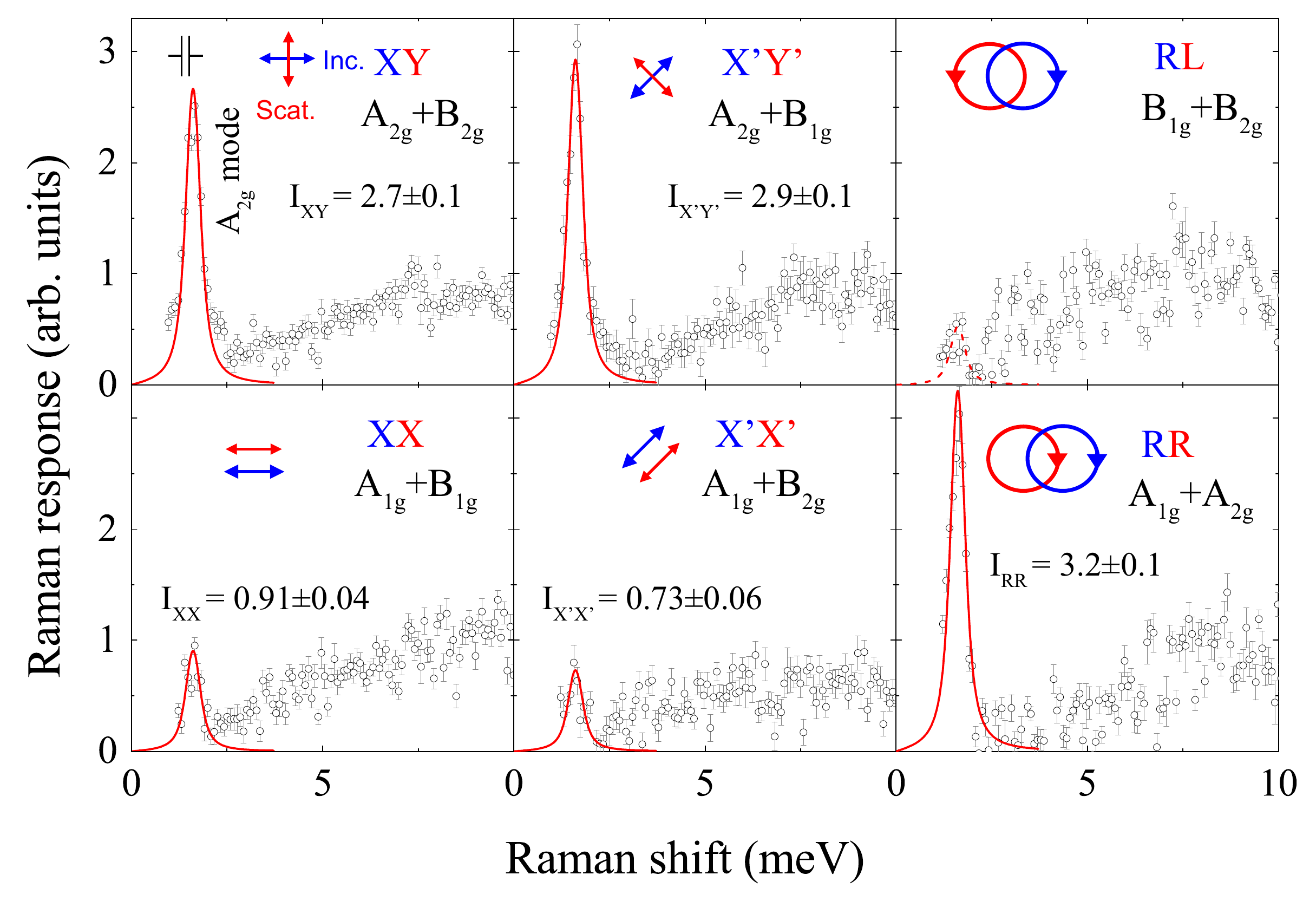}
\end{center}
\caption{
The Raman response in six proper scattering geometries at 7\,K.
The arrows in each panel show the linear or circular polarizations 
for incident (blue) and scattered (red) light.
The six proper scattering geometries are denoted as 
$\mathbf{e}_s\mathbf{e}_i=$XX, XY, X$^\prime$X$^\prime$, X$^\prime$Y$^\prime$, RR and RL, 
with $\mathbf{e}_i$ being the direction vector for incident light polarization, and
$\mathbf{e}_s$ being the scattered light polarization.
X=[100], Y=[010] are aligned along crystallographic axes,
X$^\prime$=[110], Y$^\prime$=[1$\bar{1}$0] are aligned 45$^\circ$ to the \textit{a}-axes,
R=(X+$i$Y)/$\sqrt{2}$ and L=(X-$i$Y)/$\sqrt{2}$ are right and left circularly polarized light, respectively\cite{SM}.
The irreducible representations for each scattering geometry are shown within the
$\mathbb{D_{4h}}$ point group.
The data are shown in black circles, 
where the error bars show one standard deviation.
The red solid lines are fits of the in-gap mode to a Lorentzian,
and the fitted intensity using the method of maximum likelihood is shown in each panel.
By decomposition, the in-gap mode intensity in each symmetry channels are:
$\mathrm{I_{A_{2g}}}=2.6\pm 0.1$, $\mathrm{I_{A_{1g}}}=0.7\pm 0.1$, 
$\mathrm{I_{B_{1g}}}=0.3\pm 0.1$, and $\mathrm{I_{B_{2g}}}=0.1\pm 0.1$.
The full width at half maximum of the in-gap mode is about 0.5\,meV at 7\,K
(deconvoluted with instrumental resolution of 0.15\,meV, shown in the XY panel).
}
\end{figure}

The same 1.6\,meV sharp resonance has also been observed by inelastic
neutron scattering at the commensurate crystal momentum, 
but only in the HO state~\cite{Bourdarot2003,Wiebe2007,Bourdarot2010}.  
The Raman measurement proves that this resonance is a long-wavelength excitation
of $A_{2g}$ character. 
The appearance of the same resonance in the
neutron scattering at different wavelength, 
corresponding to the \textit{c}-axis lattice constant, 
requires HO to be a staggered alternating electronic order in $c$ direction.  
Such order with alternating left and right handed states at the uranium sites 
for neighboring basal planes, 
has no modulation of charge or spin, and does not couple to tetragonal lattice,
hence it is hidden to all probes but scattering in $A_{2g}$ symmetry.  
We reveal this hidden order to be a chirality density wave depicted in figure~1D.

The chirality density wave doubles the translational periodicity of
the paramagnetic phase, hence it folds the electronic Brillouin zone,
as recently observed by angle-resolved photoemission spectroscopy\cite{Yoshida2013}. 
It also gives rise to an energy gap,
as previously observed in optics\cite{Bonn1988,Hall2012,Guo2012} and
tunneling experiments\cite{Aynajian2010,Schmidt2010}, 
and shown in figure~4 to originate in expelling the continuum of $A_{2g}$ excitations.
The sharp (0.3\,meV) resonance is explained by excitation from the ground state, 
which posses chirality density wave staggering $\ket{\mathbb{0}_\text{HO}^\text{L}}$ 
and $\ket{\mathbb{0}_\text{HO}^\text{R}}$, 
to the excited state depicted in figure~1D, 
which staggers $\ket{\mathbb{0}_\text{HO}^\text{L}}$ and $\ket{\mathbb{1}_\text{HO}}$~\cite{SM}.

A local order parameter of primary $A_{2g}$ symmetry, breaking
vertical/diagonal reflections, with subdominant $B_{1g}$ component,
breaking four-fold rotational symmetry, can be expressed in terms of the composite
hexadecapole local order parameter of the form:
\vspace{-0.05 in}
$$
\pm V[(J_x-J_y)(J_x+J_y)(J_x J_y+J_y J_x)+(J_x J_y+J_y J_x)(J_x+J_y)(J_x-J_y)]
$$
where $J_x$, $J_y$ are in-plane angular momentum operators\cite{Haule2009,SM}.
A spatial order alternating the sign of this hexadecapole for neighboring 
basal planes is the chirality density wave (see Fig.~1D)
that consistently explains the HO phenomena as it is 
observed by Raman and neutron scattering \cite{Broholm1991,Bourdarot2003,Wiebe2007,Bourdarot2010},
magnetic torque\cite{Okazaki2011}, X-ray diffraction\cite{Tonegawa2014},
and other data\cite{Bonn1988,Hall2012,Guo2012,Yoshida2013,Mydosh2011}.
Our finding is a new example of exotic electronic ordering, 
emerging from strong interaction among \textit{f} electrons, 
which should be a more generic phenomenon relevant to other intermetallic compounds.

\begin{scilastnote}
\item\textbf{Acknowledgments}
We thank J. Buhot, P. Chandra, P. Coleman, G. Kotliar, M.-A.
M\'{e}asson, D.K. Morr, L. Pascut, A. Sacuto and J. Thompson for discussions. 
G.B. and V.K.T. acknowledge support from the US Department of Energy,
Office of Basic Energy Sciences, Division of Materials Sciences and
Engineering under Award DE-SC0005463. 
H.-H.K. acknowledges support from the National Science Foundation
under Award NSF DMR-1104884. 
K.H. acknowledges support by NSF Career DMR-1405303. 
W.-L. Z. acknowledges support by ICAM (NSF-IMI grant DMR-0844115).
Work at Los Alamos National Laboratory was performed under the
auspices of the US Department of Energy, Office of Basic Energy Sciences,
Division of Materials Sciences and Engineering. \\
\end{scilastnote}

\end{document}